\documentstyle[aps,prd,preprint]{revtex}
\newcommand{\beq}{\begin{equation}}
\newcommand{\eeq}{\end{equation}}
\def\beqa{\begin{eqnarray}}
\def\eeqa{\end{eqnarray}}
\def\p{\partial}

\def\lap{\lower.5ex\hbox{$\; \buildrel < \over \sim \;$}}
\def\gap{\lower.5ex\hbox{$\; \buildrel > \over \sim \;$}}

\def\vp{\varphi}

\input{psfig.sty}

\begin{document}
\title{Predictability crisis in inflationary cosmology and its resolution}
\author{Vitaly Vanchurin, Alexander Vilenkin\\
{\em Institute of Cosmology, Department of Physics,
Tufts University, Medford MA 02155, USA}\\
and\\
Serge Winitzki\\
{\em DAMTP, University of Cambridge, Cambridge CB3 9EW, UK}
}

%\date{\today}
\maketitle
\bigskip
{\it In memory of Chandra Pathinayake}

\tightenlines

\begin{abstract}

Models of inflationary cosmology can lead to variation of observable
parameters (``constants of Nature'') on extremely large scales.  The
question of making probabilistic predictions for today's observables in
such models has been investigated in the literature.  Because of the
infinite thermalized volume resulting from eternal inflation, it has
proven difficult to obtain a meaningful and unambiguous probability
distribution for observables, in particular due to the gauge
dependence.  In the present paper, we further develop the
gauge-invariant procedure proposed in a previous work for models with a
continuous variation of ``constants''.  The recipe uses an unbiased
selection of a connected piece of the thermalized volume as sample for
the probability distribution.  To implement the procedure numerically,
we develop two methods applicable to a reasonably wide class of models:
one based on the Fokker-Planck equation of stochastic inflation, and
the other based on direct simulation of inflationary spacetime.  We
present and compare results obtained using
these methods.

\end{abstract}

\section{Introduction}

The parameters we call ``constants of Nature'' may in fact be
variables related to some slowly varying fields.  For example, what we
perceive as a cosmological constant could be a potential $U(\chi)$ of
some field $\chi(x)$.  If this potential is very flat, so that the
evolution of $\chi$ is much slower than the Hubble expansion, then
observations will not distinguish between $U(\chi)$ and a true
cosmological constant.  Observers in different parts of the universe
could then measure different values of $U(\chi)$.

Spatial variation of the fields $\chi_a$ associated with the
``constants'' can naturally arise in the framework of inflationary
cosmology \cite{1}.  The dynamics of light scalar fields
during inflation are strongly influenced by quantum fluctuations, so
different regions of the universe thermalize with different values of
$\chi_a$.  An intriguing question is whether or not we can predict
the values of the ``constants'' we are most likely to observe.

It is important to realize that this question arises not only in some
exotic models especially designed to have variable constants.  On the
contrary, we have to face it in {\it all} inflationary models attempting
to predict the spectrum of cosmological density fluctuations.  The
density fluctuation $\delta\rho/\rho(l)$ is determined by the quantum
fluctuation $\delta\phi(l)$ of the inflaton field $\phi$ at the time
when the corresponding length scale $l$ crossed the horizon.
Different realizations of random fluctuations $\delta\phi(l)$ result
in different density fluctuation spectra in widely separated parts of
the universe.  [In this case, the fluctuations $\delta\phi(l)$ on
different scales play the role of the fields $\chi_a$.]  The question
is then what spectrum $\delta\rho/\rho(l)$ we are most likely to
observe in our neighborhood?  In more general terms, we would like to
determine the probability distribution ${\cal P}(\chi)$ for the fields
$\chi_a$ associated with variable constants.

The inflationary scenario implies a very large universe inhabited by
numerous civilizations that will measure different values of
$\chi_a$.  We can define the probability ${\cal
P}(\chi)d\chi_1 ... d\chi_n$ for $\chi_a$ to be in
the intervals $d\chi_a$ as being proportional to the number of
civilizations which will measure $\chi_a$ in that interval
\cite{7}.  This includes all present, past and future civilizations; in
other words, it is the number of civilizations throughout the entire
spacetime, rather than at a particular moment of time.  Assuming that
we are a typical civilization, we can expect to observe $\chi_a$ near
the maximum of ${\cal P}(\chi)$ \cite{8}.  The assumption of being
typical has been called the ``principle of mediocrity'' in Ref.\cite{7}.

An immediate objection to this approach is that we are ignorant about
the origin of life, let alone intelligence, and therefore the number
of
civilizations cannot be calculated. However, the approach can still be
used to find the probability distribution for parameters which do not
affect
the physical processes involved in the evolution of life. The
cosmological constant $\Lambda$, the density parameter $\Omega$ and
the
amplitude of density fluctuations $Q$ are examples of such
parameters. We shall assume that our fields $\chi_a$ belong to this
category. The
probability for a civilization to evolve on a suitable planet is then
independent of $\chi_a$, and instead of the number of civilizations we
can use
the number of habitable planets or, as a rough approximation, the
number of galaxies.  [We are assuming that civilizations can exist only
for a finite period of time after inflation, either because the stars
run out of nuclear fuel and die, or because protons eventually decay.
A galaxy then gives rise to a finite average number of civilizations.]
Thus, we can write
\beq
{\cal P}(\chi)d^n\chi\propto d{\cal N},
\label{1}
\eeq
where $d{\cal N}$ is the number of galaxies that are going to be
formed in regions where $\chi_a$ take values in the intervals
$d\chi_a$.

The meaning of Eq.~(\ref{1}) is unambiguous in models where the total
number of galaxies in the universe is finite.  Otherwise, one has to
introduce some cutoff and define the ratio of probabilities for the
intervals $d^n\chi^{(1)}$ and $d^n\chi^{(2)}$ as the ratio of the galaxy
numbers $d{\cal N}^{(1)}/d{\cal N}^{(2)}$ in the limit when the cutoff
is removed.
The situation is relatively straightforward in the case of an infinite
universe which is
more or less homogeneous on very large scales.
%\footnote{This situation
%does not naturally arise in inflationary cosmology, and we mention it
%here only as an illustration.}
One can then evaluate the ratio $d{\cal N}^{(1)}/d{\cal N}^{(2)}$ in a
large comoving volume ${\cal V}$ and then take the limit as ${\cal
V}\to\infty$.  The result is expected to be independent of the
limiting procedure; for example, it should not depend on the shape of
the volume ${\cal V}$.  (It is assumed, however, that the volume
selection is unbiased, that is, that the volume ${\cal V}$ is not
carved to favor some values of $\chi_a$ at the expense of other
values.)

The problem is much more serious in models of eternal inflation where
the universe
is never completely thermalized \cite{2,11}.
In such models, the volumes of both
inflating and thermalized regions grow exponentially with time and
the number of galaxies grows without bound, even in a region of a
finite comoving size.  One can deal with this problem by
introducing a time cutoff and including only regions that thermalized
prior to some moment of time $t_c$, with the limit $t_c\to\infty$ at
the end.  One finds, however, that the resulting
probability distributions are extremely sensitive to the choice of the
time coordinate $t$ \cite{4}. This applies in particular to the
calculations of the density fluctuation spectra.  Having chosen $t$ to
be the proper time along the world lines of comoving observers,
Linde, Linde and Mezhlumian \cite{15} found that a typical
observer could find herself near the center of a very deep minimum of
the density field.  On the other hand, if one uses the expansion
factor along the trajectories as the time coordinate, one recovers the
standard results \cite{14}.
Coordinates in General Relativity are
arbitrary labels, and such gauge dependence of the results
casts doubt on any conclusion reached using this approach.

An alternative procedure, suggested in \cite{9}, is to introduce a
$\chi$-dependent cutoff at the time $t_c(\chi)$, when all but a small
fraction $\epsilon$ of the comoving volume destined to thermalize with
$\chi_a$ in the intervals $d\chi_a$ has thermalized. The limit
$\epsilon
\to 0$ is taken after calculating the probabilities. It was shown in
\cite{9,VW}  that the resulting probability distribution is
essentially insensitive to
the choice of time parametrization. However, the same problem appears
in a different guise. Linde and Mezhlumian \cite{10} have found a
family of gauge-invariant cutoff procedures which includes the
$\epsilon$-procedure described above.  All these procedures give vastly
different results for the probability distributions.

In the face of these difficulties, serious doubts have been expressed
that a meaningful definition of probabilities in an eternally
inflating universe
is even in principle possible \cite{4,GBL,10}.  Once again, this
problem cannot be brushed aside as ``exotica'' because inflation is
eternal in practically all models suggested so far.  Eternal inflation
is generic; it is finite inflation that is exotic.  We believe,
therefore, that the situation deserves to be called the
``predictability crisis in inflationary cosmology''.

In this paper we are going to analyze the resolution of the crisis
suggested by one of us in Ref.\cite{AV98}.
We begin, in Section II, by discussing the spacetime structure of an
eternally inflating universe.  This will help us clarify the origin
of the problem and will motivate its resolution proposed in Section III.
In the following Sections IV -- V we develop methods for
calculating probabilities based on the prescription introduced in
Section III.  The first method uses the Fokker-Planck equation of
stochastic inflation.  In Section IV we derive the appropriate form of
the equation, discuss the initial conditions, and work out some
examples.  The second method uses numerical simulations of eternal
inflation.  In Section V we describe our simulations and compare them
to previous work.  We
also compare the results obtained from the simulations with those
obtained using the Fokker-Planck equation.  Our conclusions are
summarized in Section VI.

\section{Spacetime structure and gauge dependence of probabilities}

An inflating Universe can be locally described using the synchronous
coordinates,
\begin{equation}
ds^2 = d\tau^2 - a^2 ({\bf x}, \tau) d{\bf x}^2 .
\label{metric}
\end{equation}
The lines of ${\bf x}={\rm const}$ in this metric are timelike geodesics
corresponding to the worldlines of co-moving observers, and the coordinate
system is well defined as long as the geodesics do not cross.  This will start
happening only after thermalization, when matter in some regions will start
collapsing as a result of gravitational instability.  Hence, the synchronous
coordinates (\ref{metric}) can be extended to the future well into the
thermalized region.

\subsection{Double-well model}

\begin{figure}
\psfig{file=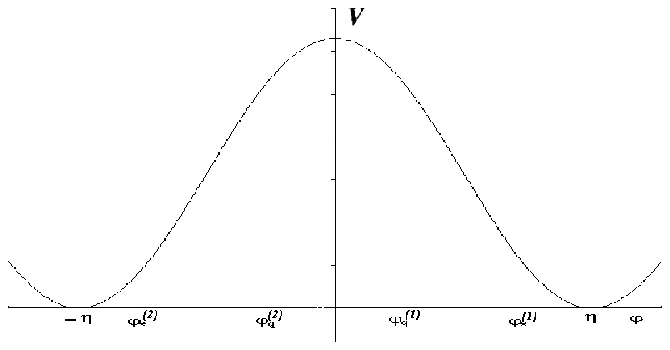,width=4.9in,angle=0}
\caption{The inflaton potential for the double-well model (\ref{v}).}
\label{fig:1}
\end{figure}

We shall first consider the simplest situation when there is a single
inflaton field $\phi$ with a double-well potential, as in Fig.~\ref{fig:1}, and
no other fields $\chi_a$.  The potential in the figure is symmetric
with respect to $\varphi\to -\varphi$, although this is not essential.
Depending on quantum fluctuations, an
inflating region with $\varphi\approx 0$ near the maximum of the
potential can thermalize in either of the two minima at
$\varphi=\pm\eta$.  The two minima may correspond to different physics,
in which case this is an example of a model
where the constants of Nature can take a discrete set of values.

In the slow-roll regime of inflation, the field $\varphi$ is a
slowly-varying function of the coordinates, so that spatial gradients of
$\varphi$ can be neglected and
\beq
{\dot \varphi}^2 \ll 2V(\varphi).
\label{sr}
\eeq
The classical evolution of the scale factor $a({\bf x},\tau)$ and of
the inflaton $\varphi ({\bf x},\tau )$ is then described by the equations
\begin{equation}
({\dot a}/a)^2 \equiv H^2 \approx 8\pi V(\varphi)/3 ,
\label{H}
\end{equation}
\begin{equation}
{\dot \varphi} \approx -V'(\varphi)/3H= -H'(\varphi)/4\pi ,
\label{phieq}
\end{equation}
where dots represent derivatives with respect to $\tau$ and we use the
Planck units, ${\hbar} = c = G = 1$.  With the
aid of Eqs.~(\ref{H}),(\ref{phieq}), the slow-roll condition (\ref{sr})
can be expressed as
\begin{equation}
H' \ll 6H .
\label{slorollcond}
\end{equation}
This condition is violated near the points $\varphi = \varphi_*^{(j)}$,
signalling the end of inflation.  The slow variation of $\varphi$ implies that
$H$ is also a slowly-varying function of ${\bf x}$ and $\tau$, and thus the
spacetime is locally close to de Sitter, with a horizon length $H^{-1}$.

Quantum fluctuations of the field $\varphi$ can be pictured as a `random walk'
[superimposed on the classical motion (\ref{phieq})] in which $\varphi$
undergoes random steps of {\it rms} magnitude $(\delta \varphi)_{\rm
rms}=H/2\pi$ per Hubble time, $\delta\tau = H^{-1}$.
The fluctuations $\delta\varphi$ are correlated on the scale of the horizon
($l \sim H^{-1}$), but correlations rapidly decay with distance
and become totally negligible on the scale of a few horizons.
% independently in each
%horizon-size region ($\ell \sim H^{-1}$) \cite{Steinh83}.
The fluctuations are
dynamically unimportant if the classical `velocity' $|{\dot \varphi}|$ is much
greater than the characteristic speed of the random walk, $(\delta\varphi)_{\rm
rms}/\delta\tau = H^2/2\pi$, which gives
\begin{equation}
H' \gg H^2 .
\label{quantcond}
\end{equation}
This condition is violated in the region $\varphi_q^{(1)} < \varphi <
\varphi_q^{(2)}$ near the top of the potential (see Fig.~\ref{fig:1}).  The dynamics of
$\varphi$ in this region is dominated by quantum fluctuations.  The
deterministic slow-roll regions are bounded by the values $\varphi_q^{(j)}$ and
$\varphi_*^{(j)}$, $j=1,2$.

\begin{figure}
\psfig{file=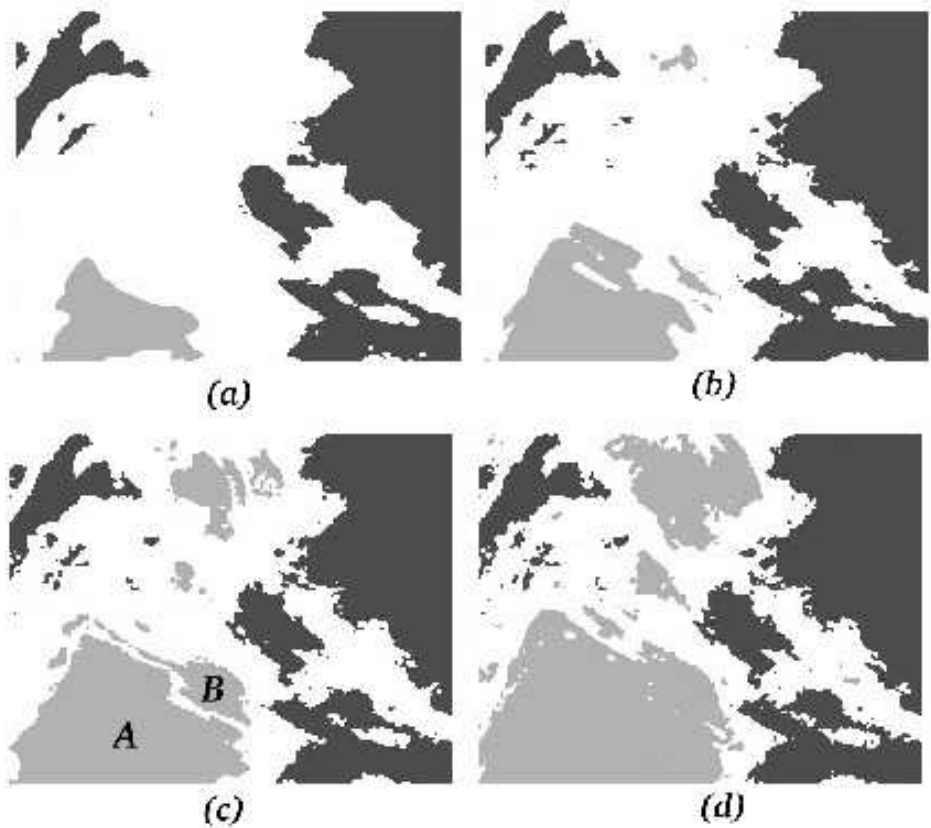,width=4.9in,angle=0}
\caption{A two-dimensional simulation for the double-well model at four 
consecutive moments of proper time: $t = 5 H_0^{-1}$(a), $t = 5.5
H_0^{-1}$(b), $t = 6 H_0^{-1}$(c), $t = 6.5 H_0^{-1}$(d). 
We evolved a comoving region of initial size 
$l=H_0^{-1}$ with the initial value of $\vp=0$ at $t=0$.  Inflating regions 
are shown white, while thermalized regions with $\vp=+\eta$ and
$\vp=-\eta$ are shown with different shades of grey. Thermalized
regions of the same type tend to join in the course of the simulation.
 For example, regions labeled $A$ and $B$ in snapshot (c) have merged
into a single region in snapshot (d).}
\label{fig:2}
\end{figure}

The random-walk picture of quantum fluctuations can be used in
numerical simulations of eternal inflation \cite{13,4}.  To illustrate
the spacetime structure of an eternally inflating universe, we have
performed simulations for several models in one and two spatial
dimensions, with and without the
additional fields $\chi_a$.  The details of the simulations are
given in Section V.  Fig.~\ref{fig:2} shows the distribution of inflating
and thermalized regions in two dimensions at four consecutive moments
of time for the
single-inflaton model with a double-well potential.  The form of the
potential is taken to be
\beq
V(\varphi)=V_0\cos^2(\alpha\varphi),
\label{v}
\eeq
where $\alpha=\pi/2\eta$; we only consider the range
$-\eta<\varphi<\eta$.  With $\alpha\lesssim 1$ and $H_0/\alpha\lesssim
1$, we have
\beq
\varphi_q^{(1,2)}\sim \pm H_0/\alpha^2,
\label{phiq}
\eeq
where $H_0=H(\varphi=0)=(8\pi V_0/3)^{1/2}$, and
\beq
\varphi_*^{(1,2)}\approx \pm(\eta-1/6).
\label{phi*}
\eeq
In the simulations we used $V_0=0.05$ and
$\alpha=1$.  Our choice of parameters in most of the simulations was dictated
by the computational constraints (see Sec.~V), so we made no attempt to
make this choice realistic.

There are two
types of thermalized regions in the model (\ref{v})
(corresponding to the two minima of the
potential at $\varphi=\pm\eta$)
which are shown with different shades of grey.  Thermalized
regions of different type can never merge; they are always separated
by inflating regions (shown in white) which can be thought of as
inflating domain walls \cite{LV}.  This is an example of
topological inflation.
Thermalized regions of the same type can also be separated by
inflating regions, but there is nothing to prevent these regions from
merging, and we can see from the sequence in Fig.~\ref{fig:2} that neighboring
regions of the same type do tend to merge.

As time goes on, more and more comoving volume gets thermalized, and
at $\tau\to\infty$ only a vanishing fraction of the initial
volume is still inflating.  However, the physical volume of the
inflating regions grows exponentially with time.  Geometrically, these
regions form a fractal of dimension $d<3$ \cite{13}.

\begin{figure}
\psfig{file=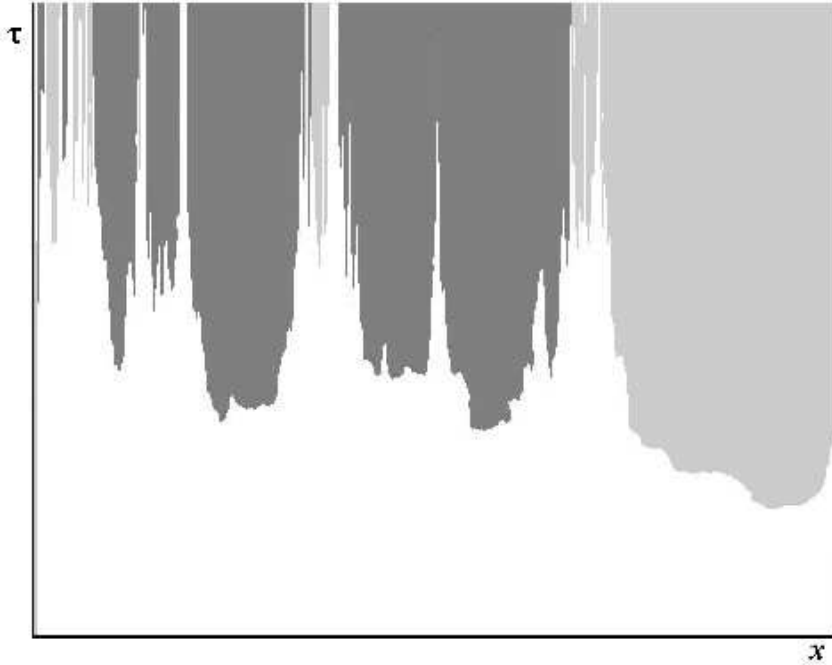,width=4.5in,angle=0}
\caption{Spacetime structure in a one-dimensional simulation for the 
double-well model.  It can be thought of as a spacetime slice through 
the ($2+1$)-dimensional simulation illustrated in Fig.~2.  Inflating 
regions are white, and thermalized regions of different type are shown 
with different shades of grey.}
\label{fig:3}
\end{figure}

A spacetime slice through the universe (in the $x-\tau$ ``plane'') is
shown in Fig.~\ref{fig:3}.  The simulation starts with a horizon-size region
and a homogeneous field $\varphi=0$ at $\tau=0$ (bottom of the figure).
We shall be particularly interested in the
thermalization hypersurfaces $\Sigma_*^{(j)}:\varphi = \varphi_*^{(j)}$
which separate inflating and thermalized spacetime regions.
We see from the figure that these surfaces tend to become asymptotically
static in the comoving coordinates as $\tau\to\infty$.
This does not mean, however, that the
surfaces become timelike at large
$\tau$.  To picture the geometry of
these surfaces, one has to keep in mind that physical lengths in an inflating
universe are related to coordinate differences by an exponentially growing
scale
factor.  As a result, the boundaries of thermalized regions (as well as all
surfaces of constant $\varphi$ in the slow-roll regime) are spacelike, and are
in fact very flat.

The spacelike character of the constant-$\varphi$ surfaces can be understood as
follows \cite{9}.  Consider the normal vector to the surfaces,
$\partial_\mu \varphi$.  We have
\begin{equation}
\partial_\mu \varphi \partial^\mu \varphi = {\dot \varphi}^2-a^{-2} ({\bf
\nabla}\varphi )^2 .
\end{equation}
The spatial gradients of $\varphi$ are caused by quantum fluctuations.  On the
scale of the horizon, the gradient is of the order $a^{-1}|{\bf \nabla}\varphi|
\sim (\delta \varphi)_{\rm rms}/H^{-1} \sim H^2 /2\pi$, and is even smaller on
larger scales.  On the other hand, from Eqs.~(\ref{phieq}),(\ref{quantcond}),
$|{\dot \varphi}| \gg H^2 /2\pi$.  Hence, $\partial_\mu \varphi \partial^\mu
\varphi > 0$, and the surfaces $\varphi = {\rm const}$ are spacelike.
%Moreover, these surfaces are locally nearly parallel to the surfaces $\tau =
%{\rm const}$ (which correspond to horizontal lines in Fig.~\ref{fig:3}).  This
%can be seen
%by considering the scalar product of the unit normals, $u^\mu = (1,0,0,0)$ and
%$n_\mu = \partial_\mu \varphi (\partial_\nu \varphi \partial^\nu
%\varphi)^{-1/2}$,
%\begin{equation}
%u^\mu n_\mu = [1 - a^{-2}({\bf \nabla}\varphi)^2 /{\dot \varphi}^2 ]^{-1/2}
%\approx 1 .
%\label{cos}
%\end{equation}

In general, thermalization hypersurfaces $\Sigma_*^{(j)}$ cannot
terminate: they must either be infinite or closed.  A closed finite
thermalization surface is possible only in a closed universe with
finite (non-eternal) inflation.  As Fig.~\ref{fig:3} illustrates, thermalization
surfaces in an eternally inflating universe extend indefinitely in the
$\tau$ direction and have, therefore, infinite volumes.  It is
easily understood that the number of disconnected thermalization
surfaces in our double-well model is also infinite.  Thermalized
regions include inflating islands which in turn include
thermalized regions of both types, and so on {\it ad infinitum}.
Regions of the same type may later merge and be, therefore, bounded
(in spacetime) by the same thermalization surface.  But clearly there
is going to be an infinite succession of nested regions of different
types which are bounded by disconnected thermalization surfaces.

For an observer in one of the thermalized spacetime regions, the surface
$\Sigma_*$ at the boundary of that region plays the role of the big
bang.  The natural choice of the time coordinate in the vicinity of that
surface is $t = \varphi$, so that the constant-$t$ surfaces are (nearly)
surfaces of constant energy density.
Since these surfaces are infinite, the observer
finds herself in an infinite thermalized universe, which is causally
disconnected from the other thermalized regions.  The situation here is similar
to that in the `open-universe' inflation, where thermalized
regions are located in the interiors of expanding bubbles and have the geometry
approximating open ($k=-1$) Robertson-Walker universes.

\begin{figure}
\psfig{file=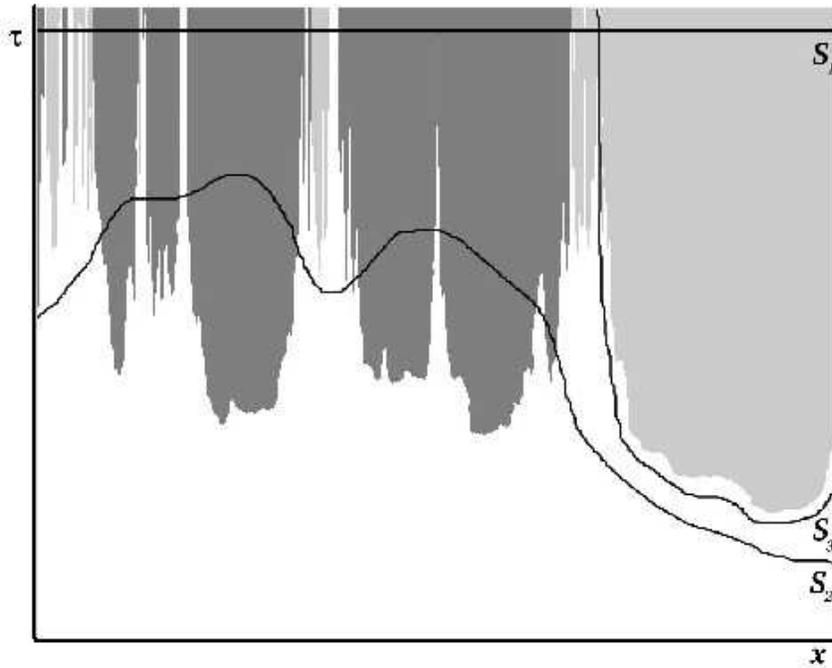,width=4.5in,angle=0}
\caption{${\cal S}_1$ is a surface of a constant proper time.  It 
crosses many thermalized regions of different types.  ${\cal S}_2$ is a 
spacelike surface which crosses regions of only one type.  ${\cal S}_3$ 
is a spacelike surface which does not cross any thermalized regions.}
\label{fig:4}
\end{figure}

It is now easy to see why the probability distributions obtained using
a cutoff at $t=t_c$ are so sensitive to the choice of the time
variable $t$.  Any spacelike surface ${\cal S}$ can be an equal-time surface
$t=t_c$ with an appropriate choice of $t$.
Depending on one's choice, the surface ${\cal S}$ may cross many thermalized
regions of different types (e.g.,
for $t=\tau$), may cross only regions of one type, or may cross no
thermalized  regions at all (say, for $t=\varphi$ with
$\varphi$ in the slow-roll range).  These possibilities are
illustrated in Fig.~\ref{fig:4} by surfaces ${\cal S}_1,{\cal S}_2$ and ${\cal
S}_3$, respectively.  If, for example, one uses the surface ${\cal
S}_2$ as the cutoff surface, one would conclude that all observers
will see the same vacuum [the same minimum of $V(\varphi)$] with
$100\%$ probability.  With a suitable choice of the surface, one can get any
result for the relative probability of the two minima.

\subsection{Two-field model}

We now consider a two-field model with a potential
\beq
V(\varphi,\chi)=V_0\cos^2(\alpha\varphi)[1+
\lambda(1+\cos\beta\chi)\sin^4\alpha\varphi].
\label{vphichi}
\eeq
The potential is symmetric with respect to $\chi\to-\chi$, and we
shall not distinguish between the values $\chi$ and $-\chi$, so the
effective range of the field $\chi$ is $0\leq\chi\leq\pi/\beta$.
%The field $\chi$ is assumed to be a cyclic variable with $\chi=0$ and
%$\chi=2\pi/\beta$ identified.
The maximum of the potential is at
$\varphi=0$ and the true vacuum is at $\varphi=\pm\eta$.  The field
$\chi$ is massless in the true vacuum, while the mass of $\varphi$ is
$\chi$-dependent,
\beq
m_\varphi^2(\chi)=2V_0\alpha^2[1+\lambda(1+\cos\beta\chi)].
\label{mphi}
\eeq
The expansion rate in the model (\ref{vphichi}) depends on both
$\varphi$ and $\chi$,
\beq
H^2(\varphi,\chi)=\frac{8\pi}{3} V(\varphi,\chi).
\label{hphichi}
\eeq
The quantum and thermalization boundaries, $\varphi_q$ and
$\varphi_*$, are also $\chi$-dependent; they are determined from
\beq
|{H'}_\varphi(\varphi_q,\chi)|\sim H^2(\varphi_q,\chi),
\eeq
\beq
|{H'}_\varphi(\varphi_*,\chi)|\sim 6H(\varphi_*,\chi).
\eeq
An important special case is when $H$ is essentially independent of
$\chi$ in the quantum regime and near thermalization.  Then
$\varphi_q$ and $\varphi_*$ are also nearly constant.  Our model
(\ref{vphichi}) belongs to this class when $\lambda$ and $\beta$ are
sufficiently small.

For small $\beta$ and $\lambda$, the $\chi$-dependence
of the potential is weak and the evolution of $\varphi$ in the
slow-roll regime is essentially the same as in the one-field model
(\ref{v}), with $\chi$ remaining nearly constant.  Different parts of
the universe will thermalize with different values of $\chi$ due to
quantum fluctuations in the random-walk regime near the top of the
potential.  This will result in spatial variation of the amplitude of
density fluctuations \cite{1},
\beq
\delta\rho/\rho(\chi)\sim 200m_\varphi(\chi).
\label{density}
\eeq
In the opposite regime, the classical evolution and quantum
fluctuations of $\chi$ are non-negligible during the slow-roll of
$\varphi$.  In this case, the density fluctuations will not be
directly related to the value of $\chi$ at thermalization, as in
Eq.~(\ref{density}).  The quantitative conditions for the two regimes
will be specified in Section IV.

In order to make the quantum dispersion
of $\chi$ during the time of the simulation comparable to the range of
$\chi$, we had to use a relatively large value of $\beta
=10$ (see Section V).  We used the same values of $V_0$ and
$\alpha$ as before and $\lambda=0.1$.  With these parameters,
one finds that $\varphi_q$ and $\varphi_*$ are essentially independent
of $\chi$ and are approximately given by
Eqs.~(\ref{phiq}),(\ref{phi*}).

\begin{figure}
\psfig{file=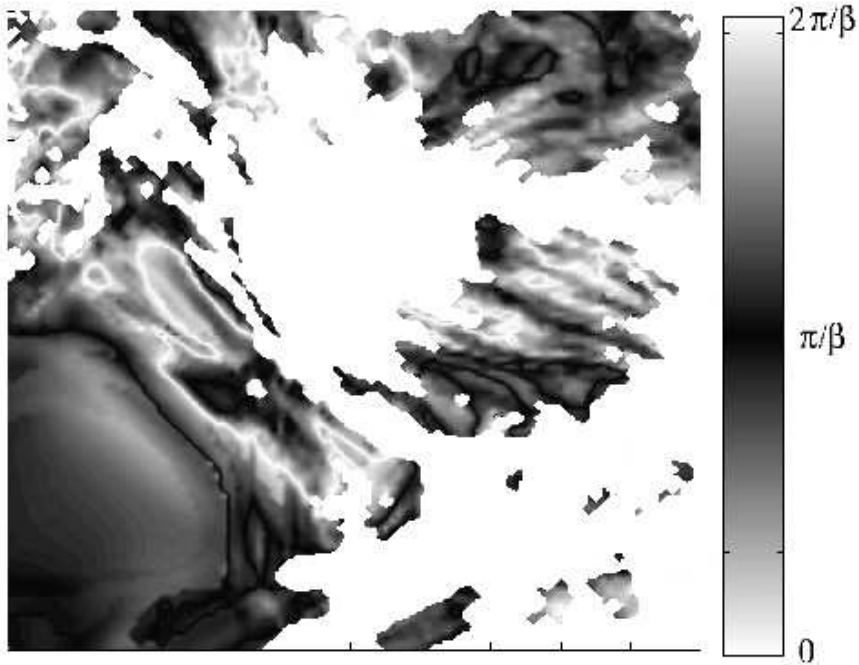,width=4.6in,angle=0}
\caption{A snapshot of a simulation for the two-field model 
(\ref{vphichi}).  Inflating regions are white, while regions that 
thermalized with different values of $\chi$ are shown with different 
shades of grey.  The shading code is indicated in the bar on the right
of the figure.}
\label{fig:5}
\end{figure}

\begin{figure}
\psfig{file=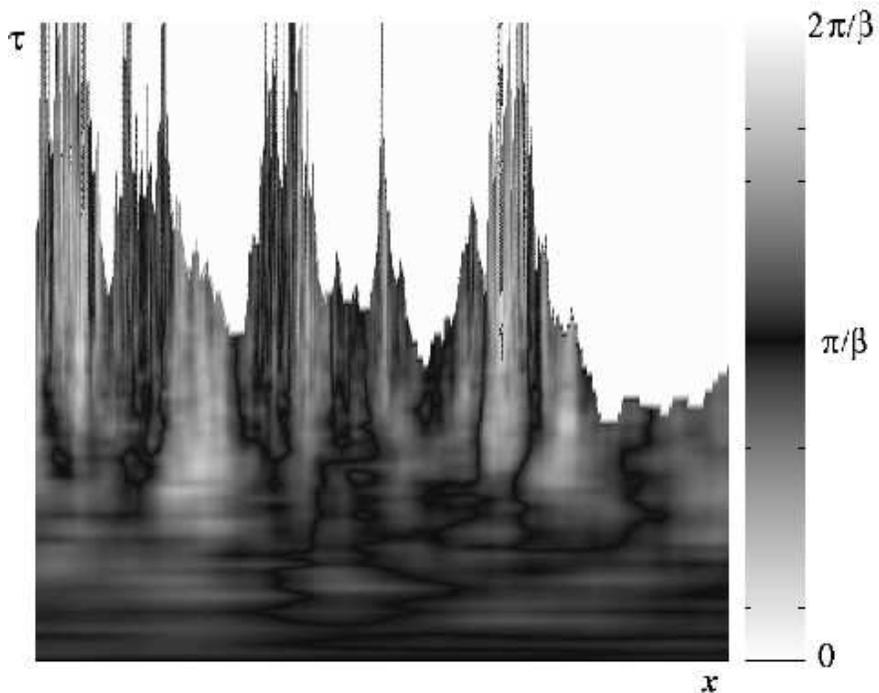,width=4.6in,angle=0}
\caption{
Spacetime structure in a one-dimensional simulation for the 
two-field model showing the evolution of a comoving region of initial 
size $l=H_0^{-1}$ with initially homogeneous $\vp=0$ and
$\chi=\pi/\beta$. 
The values of the field $\chi$ are indicated throughout the inflating region 
using the same shading code as in Fig.~\ref{fig:5}.  The thermalized 
regions are left white.}
\label{fig:6}
\end{figure}

The spacetime structure of the universe in this model is very similar
to that in the one-field model (\ref{v}).  There is an infinite number
of infinite thermalization surfaces, with the field $\chi$ varying
continuously along these surfaces.  Fig.~\ref{fig:5} gives a snapshot of a
simulation where different values of $\chi$ at thermalization are
shown with different shades of grey, from light grey for $\chi=0$ to
dark grey for $\chi=\pi/\beta$.
%Since $\chi=0$ and $\chi=2\pi/\beta$
%are identified, we have discontinuous changes of shading along the
%lines where $\chi$ takes these values.  There are, of course, no real
%physical discontinuities.
Inflating regions are white as
before, and here we do not distinguish between regions thermalizing at
$\varphi=+\eta$ and $\varphi=-\eta$, assuming that they have identical
physics.  A spacetime slice through this
simulation is shown in Fig.~\ref{fig:6}, where now the values of $\chi$ are
indicated throughout the inflating region using the same shading code,
while thermalized regions are white.

Let us now try to understand the gauge dependence of probabilities
obtained by a constant-$t$ cutoff in this model.  The simple
explanation that we gave for the one-field model, that constant-$t$
surfaces can be chosen so that they cross one type of thermalization
surfaces and avoid the other, does not apply here.  The field $\chi$
takes all its possible values on each thermalization surface
$\Sigma_*$, and the distributions of $\chi$ on different surfaces
should be statistically equivalent.  However, when we change the time variable,
say, from $t=\tau$ to $t=a$, the shape of the cutoff surface $t=t_c$
is also changed, and it is important to note that the direction in
which this surface is deformed is correlated with the local value of
$\chi$.  The scale-factor time $a$ is greater in regions where the
expansion rate $H$ is higher, and since the expansion rate is
$\chi$-dependent, the cutoff surface deformation will tend to favor
regions with $\chi$ corresponding to low values of $H$ and will tend
to exclude regions where $\chi$ corresponds to high values of $H$.
This is the origin of the gauge dependence of the cutoff procedure.

One might have thought that the deformation of the cutoff surface due
to the change of the time variable should affect only the boundaries
of the thermalized regions included in the calculation of probabilities
and should therefore have little effect in the limit $t_c\to\infty$
when the volume of the thermalized regions gets very large.  However,
the volume of thermalized regions in an eternally inflating universe
grows exponentially with time, and at any time
the total thermalized volume is dominated by the newly thermalized regions
which stopped inflating within the last few Hubble
times.\footnote{Remember that
Figs.~5,6 show the comoving volume distributions of $\chi$.  The
physical volumes are related to the co-moving volumes by the factor
$a^3({\bf x},t)$, so the volumes thermalizing at later times are
exponentially enhanced.  Moreover, as Figs.~5,6 indicate,
a fixed-time cutoff tends to produce a multitude of disconnected thermalized
regions.  Their total volume is dominated by small, newly formed
regions.}  Hence, most of the thermalized volume is always in the
vicinity of the cutoff surface $t=t_c$, and it is not surprising that
the probability distribution ${\cal P}(\chi)$ is sensitive to the
choice of the time variable $t$ used to implement the cutoff.

The spacetime structure may be different for other models of eternal
inflation, particularly for topological inflation \cite{LV} with a 
different
topology.  For example, in the case of inflating monopoles one can
expect inflating regions to be localized and to be surrounded by
thermalized regions.  It is conceivable that in this case there is a
single connected thermalized region of spacetime which is separated
from the inflating regions by a single thermalization surface
$\Sigma_*$.  Despite this possible difference, we still expect that
the constant-time cutoff procedure will still give gauge-dependent
results, for the same reasons as in the two-field model discussed
above.

\section{The proposal}

We now review the resolution of the gauge dependence problem proposed in
Ref.\cite{AV98}.  Let us first assume that inflating and thermalized
regions of spacetime are separated by a single thermalization surface
$\Sigma_*$.  The problem
with the constant-time cutoff procedures is that they cut the surface
$\Sigma_*$ in a biased way, favoring certain values of $\chi$ and
disfavoring other values.  We thus need an unbiasly selected portion
of $\Sigma_*$.

It appears that the simplest strategy is to use a ``spherical'' cutoff.
Choose an arbitrary point $P$ on $\Sigma_*$.  Define a sphere of
radius $R$ to include all points $Q$ whose geodesic
distance\footnote{If there is more than one
geodesic connecting $P$ and $Q$, $d(Q,P)$ is
defined as the smallest of all the geodesic distances.}  from $P$
along $\Sigma_*$ is $d(Q,P)\leq R$.  We can use
Eq.~(\ref{1}) to evaluate the probability distribution ${\cal P}(\chi)$
in a spherical volume of radius $R_c$ and then let $R_c\to\infty$.  If
the fields $\chi_a$ vary in a finite range, they will run through all
of their values many times in a spherical volume of sufficiently large
radius.  We expect, therefore, that the distribution ${\cal P}(\chi)$
will rapidly converge  as the cutoff radius $R_c$ is increased.  We
expect also that the resulting distribution will be independent of the
choice of point $P$ which serves as the center of the sphere.

The same procedure can be used for fields with an infinite range of
variation, provided that the probability distributions for $\chi_a$
are concentrated within a finite range, with a negligible probability
of finding $\chi_a$ very far away from that range.

Suppose now that there is an infinite number of disconnected thermalization
surfaces, as in the two-field model of Section II.  We can then pick
an arbitrary connected component of $\Sigma_*$ and apply the spherical
cutoff prescription described above.  Since the inflationary dynamics
of the fields $\chi_a$ have a stochastic nature, the distributions of
$\chi_a$ on different connected components of $\Sigma_*$ should be
statistically equivalent, and the resulting probability distribution
${\cal P}(\chi)$ should be the same for all components.

An even more complicated spacetime structure is found
in models with a non-vanishing cosmological constant.  In such models,
regions of true vacuum in the post-inflationary universe may fluctuate
back to the quantum diffusion range of the inflaton potential.
Every such region will serve as a seed for a new eternally inflating
domain whose internal structure will resemble that shown in Fig.~\ref{fig:5}.
Thermalized regions formed in this domain will in turn produce new
inflating seeds, etc. \cite{GV}.  In such a ``recycling'' universe,
there is an infinite number of thermalization surfaces to the future
of any given
thermalization surface.  But once again, we expect them all to be
statistically equivalent, so the prescription for probabilities
remains the same.

We note that this prescription cannot be applied to models where the
potential has a discrete set of minima, as in the double-well model of
Section II.  We can introduce a discrete variable $\chi$ labelling
different minima.  Each connected component of the thermalization
surface $\Sigma_*$ is then characterized by a single value of $\chi$,
and it is clear that the probability distribution for $\chi$ cannot be
determined by studying one such component.\footnote{We assume that the
discrete vacua cannot be connected by non-inflating domain walls.}
This may indicate that no meaningful probability distribution can be
defined for a discrete variable in an eternally inflating universe.
%We shall return to this issue in Section VI.  Until then we shall
%concentrate on continuous variables.

Having specified the cutoff procedure, the problem of calculating ${\cal
P}(\chi)$ can now be split into two parts.  The number of galaxies
$d{\cal N}(\chi)$ in Eq.~(\ref{1})
is proportional to the volume of the comoving
regions where $\chi_a$ take specified values and to the density of
galaxies in those regions.  The volumes and the densities can be
evaluated at any time.  Their product should be independent of the
choice of this reference time, as long as we include both galaxies
that were formed in the past and those that are going to be formed in
the future.
It will be convenient to evaluate the volumes and the densities
at the time when inflation ends and vacuum energy thermalizes, that
is, on the thermalization surface $\Sigma_*$.  Then we
can write
\begin{equation}
{\cal P}(\chi) \propto \nu(\chi){\cal P}_*(\chi).
\label{1a}
\end{equation}
Here, ${\cal P}_*(\chi)d^n\chi$ is proportional to the volume of
thermalized regions where $\chi_a$ take values in the intervals
$d\chi_a$, and
$\nu(\chi)$ is the number of galaxies that form per unit thermalized
volume with cosmological parameters specified by the values of
$\chi_a$.  The calculation of $\nu(\chi)$ is a standard astrophysical
problem
which is completely unrelated to the calculation of the volume factor
${\cal P}_*(\chi)$. Our focus in this paper will be on the volume
factor.  In the following sections we shall discuss some methods
that can be used to calculate ${\cal P}_*(\chi)$.

\section{The Fokker-Planck equation}

\subsection{Derivation of the Fokker-Planck equation}

A statistical description of fluctuating quantum fields in an eternally
inflating universe can be given in terms of the probability
distribution on equal time surfaces, $P(\varphi,\chi,t)$, which
satisfies the Fokker-Planck equation.  This description was suggested
in Ref.\cite{2}, where the idea of eternal inflation was first
introduced, and was further developed in Refs.\cite{3,GLM,Nambu,Mijic,4,SalopekBond}.
Although this is an elegant formalism, previous attempts to use it for
the calculation of probabilities gave ambiguous gauge-dependent
answers \cite{4,5,11}.  Here we suggest a version of the Fokker-Planck
formalism which allows an unambiguous calculation of ${\cal
P}_*(\chi)$ in a wide class of models.

The idea is that for $\varphi\gg\varphi_q$, when quantum fluctuations
of $\varphi$ (but not necessarily of $\chi$) are negligible compared
to its classical motion, the evolution of $\varphi$ at any given co-moving
point is essentially monotonic in proper time and we can use
$\varphi$ as a new time variable, $t=\varphi$. Let us first assume that
the $\chi$-dependence of the expansion rate $H$ is insignificant
near thermalization, so that $\varphi_*$ is also nearly independent of
$\chi$.  Then the thermalization boundary
$\varphi=\varphi_*$ is a constant-``time'' surface and
the probability distribution ${\cal P}(\chi)$ is given simply by
$P(\varphi_*,\chi)$.  We shall now derive the Fokker-Planck equation
for $P(\varphi,\chi)$.
That equation will be applicable in the regime $\varphi \gg\varphi_q$ where
the evolution of $\varphi$ is essentially classical (but without such
restrictions on $\chi$), and will have to be complemented by appropriate
boundary conditions (see Section IV, B).

The classical drift velocity of $\chi$ in time $t=\varphi$ can be
written as
\beq
v_\chi\equiv{d\chi\over{d\varphi}}={{\dot\chi}\over{\dot\varphi}}
={{H'}_\chi \over{{H'}_\varphi}},
\label{eq:v-red}
\eeq
where dots indicate derivatives with respect to the proper time
$\tau$, $H(\varphi,\chi)$ is given by Eq.~(\ref{hphichi}) and
${H'}_\chi=\partial H/\partial\chi$, etc.  The quantum dispersion
during the ``time'' increment $d\varphi$ is
\beq
(\delta\chi_q)^2={H^3\over{4\pi^2}}d\tau=-{H^3\over{\pi{H'}_\varphi}}
d\varphi \equiv 2D(\varphi,\chi)d\varphi,
\label{eq:D-red}
\eeq
where $D(\varphi,\chi)$ is the diffusion coefficient.  Finally, the
rate of the cosmological expansion is
\beq
{1\over{a}}{\partial a\over{\partial\varphi}}={H\over{\dot\varphi}}
=-4\pi{H\over{{H'}_\varphi}}.
\eeq
The Fokker-Planck (FP) equation for $P(\varphi,\chi)$ can now be
written as
\beq
{\p P\over{\p\varphi}}={\p^2\over{\p\chi^2}}(DP)-{\p\over{\p\chi}}
(v_\chi P)-4\pi d{H\over{{H'}_\varphi}}P,
\label{FP}
\eeq
where $d$ is the number of spatial dimensions.
The first and second term on the right-hand side of (\ref{FP})
describe the quantum diffusion and the classical drift of $\chi$,
respectively, and the last term accounts for the expansion of the
universe.  Eq.~(\ref{FP}) can also be derived directly from the FP
equation for the two-field distribution function $P(\varphi,\chi,t)$;
this is done in Appendix A.
We should note that due to the growth of physical volume, the
distribution $P(\varphi,\chi)$, just like $P(\varphi,\chi,t)$, is not
normalized. The physical probability distribution for $\chi$ is given
by $P(\varphi_{*},\chi)$ after an appropriate normalization of that
function over the range of $\chi$.

Starobinsky has pointed out \cite{3} that the diffusion term of the FP
equation suffers from the ambiguity in the ordering of non-commuting
factors $\p/\p\chi$ and $D(\varphi,\chi)$.  For example, one could
write this term as
\beq
{\p\over{\p\chi}}\left[ D^{1/2+\gamma}{\p\over{\p\chi}}\left(D^{1/2
-\gamma} P\right)\right].
\eeq
However, it has been recently shown \cite{AV99} that the so-called Ito factor
ordering, $\gamma=-1/2$, has significant advantages compared to other
choices.  Hence, we used $\gamma=-1/2$ in Eq.~(\ref{FP}).

It is also possible to relax the assumption that the thermalization surface \(
\phi =\phi _{*} \) is independent of \( \chi  \).  Usually the endpoint \( \phi
_{*} \) is defined only approximately and one could redefine its exact position
(within the range where fluctuations of both \( \chi \) and \( \phi \) are
negligible) to remove the dependence of \( \phi_{*} \) on \( \chi \), without
changing the physical results. In case such redefinition is impossible, the
formalism can be extended to give the distribution of \( \chi  \) along an
arbitrary boundary line specified by a function \( \phi _{*}\left( \chi \right)
\).  To implement this, one formally puts the diffusion coefficient \( D\left(
\phi ,\chi \right)  \)  and the drift velocity \( v_{\chi } \) to zero everywhere
in the domain \( \phi >\phi _{*}\left( \chi \right)  \)  and solves the modified
FP equation in the entire range of \( \phi  \) and \( \chi  \) up to a suitable
value \( \phi =\phi _{\max } \) which should lie beyond the boundary \( \phi
_{*}\left( \chi \right)  \) for all \( \chi  \).  The solution of the new FP
equation adequately describes fluctuations of \( \chi  \)  up to the boundary,
while beyond the boundary the distribution is kept unchanged  for a given \(
\chi  \). The restriction of this solution to \( \phi =\phi _{\max } \)  (clearly
independent of the choice of \( \phi _{\max } \)) gives the desired probability
distribution \( \mathcal{P}_{*}\left( \chi \right)  \). We shall therefore assume
below that the boundary \( \phi _{*} \) is independent of \( \chi  \).

\subsection{Initial conditions}

To solve the FP equation (\ref{FP}) we have to specify the initial
distribution $P(\varphi_0,\chi)$ at some $\varphi=\varphi_0$ in the
deterministic slow-roll range, $\varphi_q(\chi)\ll\varphi_0
<\varphi_*(\chi)$.
We do not know how to do this in the general
case.  However, a simple initial distribution can be specified in
models where the potential $V(\varphi,\chi)$ is very symmetric near
the top, so that it is essentially independent of $\chi$ in the
interval $0\leq\varphi<\varphi_1$ which includes the diffusion range
and part of the slow-roll range ($\varphi_q\ll\varphi_1<\varphi_*$).
We can then choose $\varphi_0$ to be in the part of the slow-roll
range where the potential is still symmetric ($\varphi_q\ll \varphi_0
<\varphi_1$).
%For definiteness we assume that $\varphi_0>0$.

Let us consider the (spacelike) surface $\Sigma_0:\varphi(x)=\varphi_0$.
In the slow-roll regime $\varphi$ rolls from smaller to larger values,
and thus the values of $\varphi$ in the past of $\Sigma_0$ are in the
range $\varphi_q\lesssim\varphi<\varphi_0$, and prior to that in the
diffusion range $-\varphi_q\lesssim\varphi\lesssim\varphi_q$.  In both
of these ranges the potential is nearly independent of $\chi$, and therefore
we expect the distribution $P(\varphi_0,\chi)$ on $\Sigma_0$ to be
flat,
\beq
P(\varphi_0,\chi)=const,
\label{ic}
\eeq
with all values of $\chi$ equally probable.
%The problem is thus
%reduced to solving Eq.~(\ref{FP}) with the initial condition (\ref{ic}).

We shall now specify the conditions of applicability of the distribution
(\ref{ic}).  Suppose the potential $V(\vp,\chi)$ can be represented as
\beq
V(\vp,\chi)=U(\vp) +{\cal V}(\vp,\chi)
\label{VUF}
\eeq
with ${\cal V}\ll U$.
The expansion rate can be written as
\beq
H(\vp,\chi)=H(\vp)+\delta H(\vp,\chi),
\label{HdH}
\eeq
where
\beq
\delta H(\vp,\chi)\approx 4\pi{\cal V}(\vp,\chi)/3H(\vp).
\label{dH}
\eeq
Let us consider the physical processes affecting the distribution
of $\chi$ and their characteristic timescales.
The $\chi$-dependent
correction to the expansion rate (\ref{dH}) will tend to distort the
distribution, producing a peak at the maximum of ${\cal V}(\vp,\chi)$
where the expansion rate is the highest.  This differential expansion
becomes important
on a timescale $\tau_{de}$ such that $\delta H\tau_{de}\sim 1$,
\beq
\tau_{de}\sim 1/\delta H.
\label{taude}
\eeq
On the other hand, the classical force will tend to drive $\chi$ to
the minimum of ${\cal V}(\vp,\chi)$ on a timescale
\beq
\tau_c\sim {\Delta_\chi\over{v_c}}\sim
4\pi{\Delta_\chi^2\over{\delta H}},
\label{tauc}
\eeq
where we have used $v_c=-\delta H'_\chi /4\pi\sim\delta
H/4\pi\Delta_\chi$ for the classical drift velocity.\footnote{Note
that $v_c$ is different from $v_\chi$ in Eq.~(\ref{v}) because the
velocity $v_\chi$ is defined with respect to ``time'' $\vp$, while
$v_c$ is defined with respect to the proper time.}
Finally, the quantum diffusion of $\chi$ will tend to keep
the distribution flat.
The typical variation of $\chi$ due to the quantum random walk
in a proper time interval $\tau$ is $(\Delta\chi)^2\sim
(H^3/4\pi^2)\tau$.  The time $\tau_q$
it takes $\chi$ to spread over its entire range $\Delta_\chi$ is
\beq
\tau_q\sim \Delta_\chi^2/H^3.
\label{tauq}
\eeq

The effect of randomization due to diffusion will prevail over the
differential expansion and classical drift, provided that
\beq
\tau_q\ll\tau_c,~\tau_{de},
\label{C1}
\eeq
or,
\beq
{\Delta_\chi^2\delta H\over{H^3}}\ll 1,~~~~~~~~{1\over{4\pi}}{\delta
H\over{H^3}} \ll 1.
\label{C1'}
\eeq

If conditions (\ref{C1'}) are satisfied at
$\vp\lesssim\vp_q$, then we expect
that   $P(\vp_q,\chi)\approx const$ at the onset of the slow roll.
At $\vp\gg\vp_q$, the inflaton $\vp$ grows monotonically with
time, causing an evolution of the expansion rate $H$ and of the
potential ${\cal V}$.  This introduces two additional timescales
\beq
\tau_H\sim H/{\dot H}\sim 4\pi H/{H'}_\vp^2
\label{tauH}
\eeq
and
\beq
\tau_{\cal V}\sim {\cal V}/{\dot{\cal V}}\sim {4\pi\over{{H'}_\vp}}
{\delta H\over{{\delta H'}_\vp}}.
\label{tauV}
\eeq
If both of these times are large compared to $\tau_q$, then
the flat distribution of $\chi$ extends into the slow roll regime
for as long as $\tau_q$ remains the smallest of all relevant
times.  In fact, the evolution of $H$ and
${\cal V}$ does not by itself distort the
flat distribution of $\chi$, and one can expect that this distribution
persists as long as the conditions (\ref{C1'}) are satisfied, even if
$\tau_q$ gets larger than $\tau_H$ or $\tau_{\cal V}$.

Another interesting case where the probability distribution for $\chi$
can be specified is
when $\tau_c$ and $\tau_q$ are small compared to all
other relevant timescales,
\beq
\tau_c,~\tau_q\ll\tau_{de},~\tau_H,~\tau_{\cal V}.
\label{condeq}
\eeq
Then the distribution $P(\vp,\chi)$ is
determined by a statistical equilibrium between the
quantum diffusion and the classical drift.
This equilibrium distribution is given by \cite{SY}
\beq
P_{eq}(\vp,\chi)\propto \exp\left(-{8\pi^2{\cal
V}(\vp,\chi)\over{3H^4(\vp)}}\right).
\label{Peq}
\eeq
It is an approximate solution of Eq.~(\ref{FP}) when the
$\vp$-dependence of $D$ and $v_\chi$ can be disregarded and the last
term in (\ref{FP}) can be dropped.  We note that it follows
from Eqs.~(\ref{tauc}),(\ref{tauq}) and (\ref{dH}) that 
\beq
{8\pi^2{\cal V}\over{3H^4}}\sim 8\pi^2{\tau_q\over{\tau_c}}.
\label{taus}
\eeq
Hence, for $8\pi^2\tau_q\ll\tau_c$ the distribution (\ref{Peq}) 
is nearly flat, as expected.
In the opposite limit, $\tau_c\ll\tau_q$, the distribution is strongly
peaked at the minimum of ${\cal V}(\vp,\chi)$.
%If ${\cal V}(\vp,\chi)\ll H^4$ for all $\chi$, then
%(\ref{Peq}) gives a flat distribution, as before.  Otherwise,
%$P(\vp,\chi)$ is suppressed for values of $\chi$ where ${\cal V}>H^4$.
%The relaxation time $\tau_{rel}$ in this case should be understood
%as the time it takes to relax to the equilibrium.  It is still given
%by Eq.~(\ref{taurel}), but now $\Delta_\chi$ is the characteristic
%range of $\chi$ which is determined from ${\cal V}(\vp,\chi)\lesssim
%H^4(\vp)$.
If the conditions of equilibrium (\ref{condeq}) are satisfied at some
$\vp_0$ in the range $\vp_q\ll\vp_0<\vp_*$, we can use
$P_{eq}(\vp_0,\chi)$ as the initial condition for the FP equation at
$\vp=\vp_0$.

\subsection{Some examples}

We shall now give some examples of numerical and analytic solutions of the FP
equation (\ref{FP}) for the two-field model (\ref{vphichi}). In this model
$U(\vp)=V_0\cos^2(\alpha\vp)$,
${\cal V}(\vp,\chi)=\lambda V_0\cos^2(\alpha\vp) \sin^4
(\alpha\vp)(1+\cos\beta\chi)$, $\Delta_\chi=\pi/\beta$, and
the conditions (\ref{C1'}) which ensure a
flat distribution at the onset of the slow roll take the form
\beq
\lambda H_0^2/8\pi\alpha^4\ll 1,~~~~~~~~~~~~~
(\pi/\beta)^2(\lambda H_0^2/\alpha^4) \ll 1.
\label{C1''}
\eeq
Here we have used $\vp\sim\vp_q\sim H_0/\alpha^2$.

\begin{figure}
\psfig{file=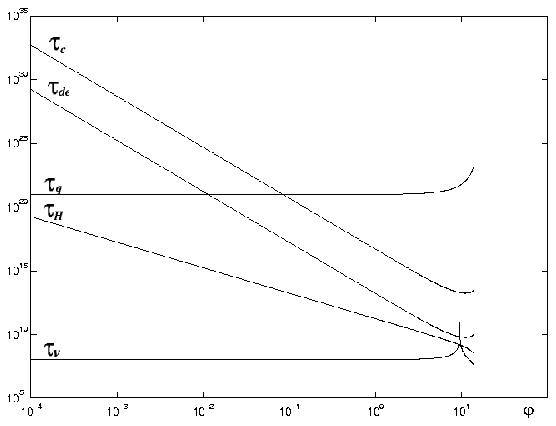,width=4.8in,angle=0}
\caption{The characteristic times as functions of $\vp$ for the model 
parameters (\ref{ex1}).}
\label{fig:7}
\end{figure}

\begin{figure}
\psfig{file=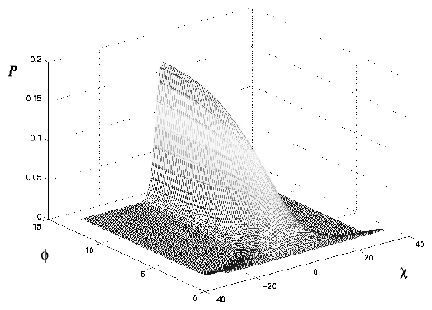,width=6in,angle=0}
\caption{The solution of the FP equation for the model parameters (\ref{ex1}).}
\label{fig:8}
\end{figure}

In our first example we used
\beq
\alpha=0.1,~~\beta=0.1,~~\lambda=0.0 1,~~H_0=10^{-6}
\label{ex1}
\eeq
and the initial
condition  $P=const$ at $\vp_0=10^{-3}$ (which is above
$\vp_q\sim 10^{-4}$).  These values of the parameters could occur in a
realistic inflationary model: Eqs.~(\ref{mphi}) and  (\ref{density}) give
$m_\phi\sim 5\times 10^{-7}$ and $\delta\rho/\rho\sim 10^{-5}$.
The characteristic times for the
parameters (\ref{ex1})
are plotted in Fig.~\ref{fig:7} as functions of $\vp$.  We see that
the conditions (\ref{C1}) are initially satisfied, and thus we are
justified to use a flat distribution as the initial condition.
The result of a numerical integration of
Eq.~(\ref{FP}) is shown in Fig.~\ref{fig:8}.  At
$\vp=\vp_*$, the distribution is peaked at $\chi=0$, indicating that
the effect of differential expansion was more important than that of
classical drift and of diffusion.  This is not surprising, since
$\tau_{de}\ll \tau_q,~\tau_c$ in most of the slow roll range.

The classical variation of $\chi$ during the slow roll of $\vp$ from some
$\vp_0>\vp_q$ to $\vp_*$ is given by
\beq
\delta\chi_c=\int_{\vp_0}^{\vp_*}{{\cal V'}_\chi\over{{U'}_\vp}}d\vp
\sim \lambda\beta/\alpha^2
\label{chic}
\eeq
and is essentially independent of $\vp_0$.  Similarly, the variation of
$\chi$ due to quantum fluctuations is
\beq
(\delta\chi_q)^2=\int{H^3\over{4\pi^2}}d\tau=-{1\over{\pi}}
\int_{\vp_0}^{\vp_*} {H^3\over{{H'}_\vp}}d\vp\approx
{H_0^2\over{\pi\alpha^2}}\ln{\sin(\alpha\vp_*)\over{\sin(\alpha\vp_0)}},
\label{chiq}
\eeq
with only a logarithmic dependence on $\vp_0$.
For the parameter values in our example,
both $\delta\chi_c$ and $\delta\chi_q$ are small compared to the
range of $\chi$, $\Delta_\chi=\pi/\beta$, and thus $\chi$ remains
nearly constant during the slow roll.

When both diffusion and classical drift are negligible,
the first two terms on the
right-hand side of the FP equation (\ref{FP}) can be dropped,
and the solution of (\ref{FP}) is
\beq
P(\vp_*,\chi)\propto\exp\left(-4\pi d\int_{\vp_o}^{\vp_*}{H(\vp,\chi)
\over{{H'}_\vp(\vp,\chi)}}d\vp\right)=e^{N(\chi)d}.
\label{oldsol}
\eeq
Here,
\beq
N(\chi)=\int Hd\tau=-4\pi\int_{\vp_0}^{\vp_*}{H\over{{H'}_\vp}}d\vp
\approx N_0+{2\pi\lambda\over{\alpha^2}}\cos(\beta\chi)
\eeq
is the number of inflationary e-foldings between $\vp_0$ and $\vp_*$,
$N_0$ is the number of e-foldings in the $\chi$-independent case
($\lambda=0$), and we have used $\sin(\alpha\vp_*)\approx 1$, which is
valid for $\alpha\lesssim 1$ [see Eq.~(\ref{phi*})].
Eq.~(\ref{oldsol}) has a simple interpretation \cite{AV98}: if $\chi$ remains
nearly constant, then the probability distributions for $\chi$ at
$\vp=\vp_0$ and $\vp=\vp_*$ are related simply by the volume expansion
factor $e^{Nd}$.  Our numerical solution (which was done for $d=3$)
agrees with the analytic solution (\ref{oldsol}) within 10\%.

\begin{figure}
\psfig{file=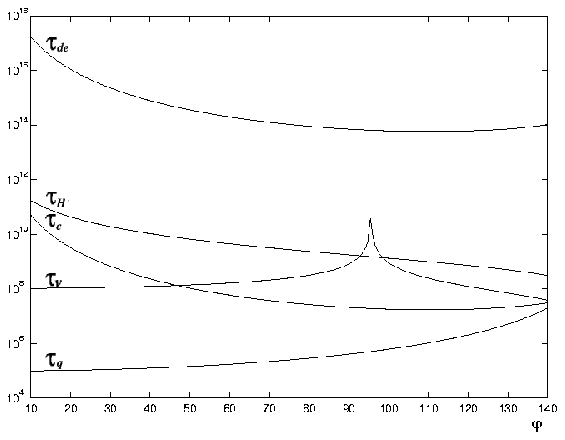,width=4.8in,angle=0}
\caption{The characteristic times as functions of $\vp$ for the model
parameters (\ref{ex2}).}
\label{fig:9}
\end{figure}

\begin{figure}
\psfig{file=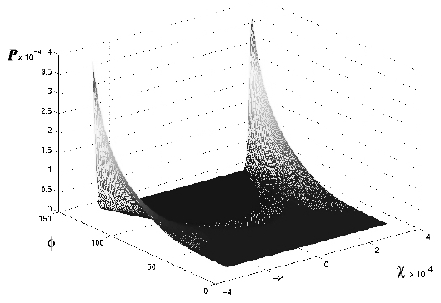,width=6in,angle=0}
\caption{The solution of the FP equation for the model parameters 
(\ref{ex2}).}
\label{fig:10}
\end{figure}

Our second example is
\beq
\alpha=0.01,~~\beta=10^4,~~\lambda=10^{-8},~~ H_0=10^{-4}
\label{ex2}
\eeq
with a flat initial condition at $\vp=10$.  The characteristic
timescales for this case are plotted in Fig.~\ref{fig:9} and the numerical
solution of the equation is shown in Fig.~\ref{fig:10}.  In this
example, $\tau_q$ 
is initially much smaller than all other characteristic times, and thus the
use of the flat initial condition is justified.  The distribution
develops a peak at $\chi=\pi/\beta$ at ``time'' $\vp\sim 60$, when
$8\pi^2\tau_q/\tau_c\sim 1$, as expected [see Eq.(\ref{taus})].  
$\tau_c$ and
$\tau_q$ remain smaller than other characteristic times throughout
the slow roll period, so one expects that the equilibrium distribution
(\ref{Peq}) should be an approximate solution of the FP equation
(\ref{FP}).  We have verified that this is indeed the case, with a
very good accuracy.  The two distributions begin to diverge as we
approach thermalization ($\vp\sim 140$), presumably because
$\tau_{\cal V}$ becomes comparable to $\tau_c$ and $\tau_q$.  However,
even at $\vp=140$ the equilibrium distribution is accurate within
$\sim$ 5\%.  We had to terminate the numerical solution
at $\vp=140$, somewhat short of $\vp_*\approx 157$, because the peak
of the distribution was getting too narrow.

\begin{figure}
\psfig{file=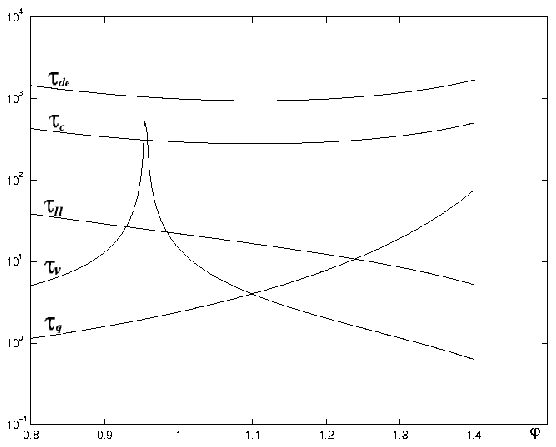,width=4.8in,angle=0}
\caption{The characteristic times as functions of $\vp$ for the model
parameters (\ref{ex3}).}
\label{fig:11}
\end{figure}

\begin{figure}
\psfig{file=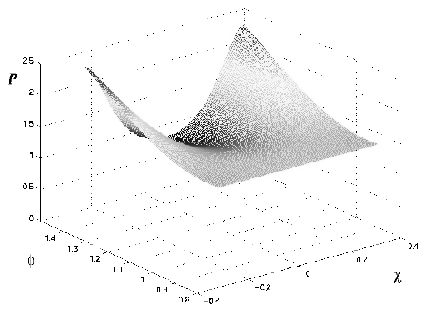,width=6in,angle=0}
\caption{The solution of the FP equation for the model parameters 
(\ref{ex3}).}
\label{fig:12}
\end{figure}

Finally, we present a numerical solution of Eq.~(\ref{FP}) for the
parameters we used in the simulations:
\beq
\alpha=1,~~\beta=10,~~\lambda=0.1,~~H_0=0.63.
\label{ex3}
\eeq
with a flat initial condition at $\vp_0=0.8$.  The simulations were
performed in ($1+1$) dimensions, so we used $d=1$.
(This choice of parameters was dictated by computational constraints;
see Section V).  The characteristic timescales for this example are
plotted in Fig.~\ref{fig:11} and the solution is shown in Fig.~\ref{fig:12}.
Once again, we see that initially
$\tau_q$ is smaller than all other timescales, and thus our choice of
the initial condition is justified.  But towards the end of the slow
roll, $\tau_c$, $\tau_{de}$ and $\tau_H$ become comparable
to $\tau_q$, within an order of magnitude.  So in this example the
shape of the final distribution at $\vp=\vp_*$ is hard to predict
without a numerical solution of the FP equation.

For the parameter values (\ref{ex3}) we have $\vp_q\sim 0.63$, which
is close to the initial value $\vp_0=0.8$.  Strictly speaking, $\vp$
cannot be used as a time variable in such proximity of $\vp_q$.
However, our solution of the FP equation is in a good agreement with
the distribution $P(\vp,\chi)$ obtained from numerical simulations
(see Section V).  This indicates that our method is robust and gives
a reasonable accuracy even for $\vp_0\sim\vp_q$.

\section{Numerical simulations}

The most direct approach to the calculation of ${\cal P}_*(\chi)$ is
to use a numerical simulation of eternally inflating spacetime.  The first
simulation of this sort \cite{13} was performed on a square lattice
representing a comoving region of initial size $H^{-1}$ in a
$(2+1)$-dimensional universe.  The initial value of the inflaton
$\varphi$ was set to $\varphi=0$ at all points of the
lattice.  The potential was assumed to be very flat in the whole range
$|\varphi|<\varphi_*$, so that $H\approx const$ throughout the
inflating region.  Then there is no classical slow roll, and the
evolution of $\varphi$ is driven entirely by quantum
fluctuations.  The fluctuations were simulated by adding random
increments to $\varphi$ independently in each horizon-size region.
After each two-folding of expansion, square regions
which had physical size $H^{-1}$ at the previous step were subdivided
into four smaller squares and a fluctuation, drawn from a Gaussian
distribution of width $(H/2\pi)(\ln 2)^{1/2}$, was added to $\varphi$
in each of these squares.  Whenever $|\varphi|$ exceeded $\varphi_*$,
the corresponding cell was marked as thermalized.

% Start of corrected text

Linde, Linde and Mezhlumian \cite{LLM,4} developed more realistic
simulations which allow for the slow roll of $\varphi$.  They
performed simulations of inflating domain walls and vortices in
$(2+1)$ dimensions.  The classical change of $\varphi$ in a small
proper time
increment $\Delta \tau$ was taken as [see Eq.~(\ref{phieq})]
\beq
\Delta\varphi=-\frac{V'(\varphi)}{3H}\Delta \tau
\label{eq:dphic}
\eeq
and quantum fluctuations were represented by sinusoidal waves
\beq
\delta\varphi({\bf x})={H\over{\sqrt{2}\pi}}\sqrt{H\Delta \tau}
\sin\left( He^{H\tau}({\bf n\cdot x}+\alpha)\right)
\label{sin}
\eeq
of wavelength comparable to the horizon, $\ell =2\pi/H$.  A phase
$\alpha$ and a unit two-dimensional vector ${\bf n}$ were chosen at
random at each step of the simulation.
The expansion rate $H$ was still treated as a constant.

LLM have also performed $(1+1)$ and $(2+1)$-dimensional simulations in
which they included the effect of variation of the expansion rate $H$
in space and time \cite{4}.  In $(1+1)$ dimensions they still used the
sinusoidal {\it ansatz} for the fluctuations, but now the amplitude and
the wavelength of the sine depended on the local value of $H(x,\tau)$. 
In $(2+1)$ dimensions with a variable $H(x,\tau)$ this {\it ansatz}
could not be used and LLM represented the fluctuations by wavelets. 

An advantage of using the sine waves is that they give a continuous
variation of $\varphi$, in contrast to the discontinuous jumps along
the boundaries of the squares in the method of Ref.~\cite{13}. However,
use of the sine functions presents two problems. Firstly, the sine
waves introduce unphysical correlations between the values of
$\delta\varphi$ at points separated by large distances\footnote{The
unphysically large long-distance correlations resulting from
Eq.~(\ref{sin}) disappear only in the limit of $\Delta \tau \rightarrow
0$. These issues will be addressed in more detail in Ref.~\cite{VW99}.
}. In this respect the simulation of Ref.~\cite{13}, which treated
$\delta\varphi$ in neighboring horizon regions as completely
uncorrelated, was more realistic. We note, however, that the
calculation of probability distribution will be unaffected by these
extra correlations, provided that the values of $\chi$ are sufficiently
well sampled throughout the volume. Also, the wavelet representation of
fluctuations, as used by LLM in $2+1$-dimensional simulations,
eliminates correlations at distances beyond the wavelet size (which was
taken to be larger than the horizon size in LLM).

Secondly, the fluctuation amplitude computed with the sine wave {\it
ansatz} is strictly bounded from above by the value it takes when the
sine function in Eq.~(\ref{sin}) is equal to one.  In reality, the
amplitude of the fluctuations is a Gaussian variable, and large values
of $\delta\varphi$ are suppressed but not strictly forbidden. LLM have
dealt with this problem by using a small timestep ($\Delta \tau \ll
H^{-1}$), effectively using a superposition of many sine waves with
random phases in each Hubble time. This approximates the desired
Gaussian distribution significantly better than a single sine wave
(although it does not eliminate unphysical correlations at large
distances).

Motivated by these considerations, we have attempted to develop
simulations that do not suffer from these problems, as described in
Sections V A.

%End

Numerical simulations of eternal inflation 
encounter significant
computational constraints, and we were able to perform simulations
only for a restricted range of parameter values.
%To illustrate the origin of the constraints, we present some rough
%estimates assuming for simplicity that $H\approx const.$
Suppose we do the simulation starting with $\vp= 0$
%and some arbitrary value of $\chi$
in a horizon-size region, $l_0=H_0^{-1}$.  After $N$
e-foldings of expansion, the size of our comoving region is
$l=H_0^{-1}e^N$.
% and the dispersion of $\chi$ due to quantum fluctuations
%is\footnote{The quantum dispersion of $\chi$ is developed mainly during the
%random-walk near the top of the potential, and the assumption
%$H\approx const$ is not unreasonable in this regime.}
%\beq
%(\Delta\chi)^2\sim (H/2\pi)^2 N=(H/2\pi)^2\ln (Hl).
%\label{deltachi}
%\eeq
The simulation has
to stop when there are still a few grid points per horizon, and
realistically the total number of points cannot exceed ${\cal
N}_{max} \sim 10^7$.  Then, in a $(d+1)$-dimensional simulation, the
number of inflationary e-foldings is
\beq
N\lesssim d^{-1}\ln{\cal N}_{max}+\ln(H_0/H_*)\sim 16/d,
\label{N}
\eeq
where we have assumed that $\ln(H_0/H_*)\sim\ln(6/\alpha)\lesssim
16/d$.

The number of inflationary e-foldings in the slow-roll range
$\vp_q<\vp <\vp_*$ is
\beq
N=-4\pi\int_{\vp_q}^{\vp_*}{H\over{{H'}_\vp}}d\vp\approx {4\pi\over
{\alpha^2}} \ln{\sin(\alpha\vp_*)\over{\sin(\alpha\vp_q)}} \approx
{4\pi\over{\alpha^2}} \ln{\alpha\over{H_0}},
\eeq
and it follows from Eq.~(\ref{N}) that we need $\alpha\gtrsim 1$.
%The typical variation of $\varphi$ due to quantum fluctuations is
%$\Delta\vp\sim (H/2\pi)\sqrt{N}$.  Requiring that $\Delta\vp$
%exceeds the width of the
%quantum diffusion range $\varphi_q$  we have, using
%Eqs.~(\ref{phiq}),(\ref{N}),
%$\alpha^2\gtrsim 2\pi/\sqrt{N}\gtrsim 1$.  (A similar constraint is
%obtained by requiring that the available number of e-foldings is
%sufficient for $\vp$ to cross the slow-roll range.)
On the other hand, $\alpha$
is bounded from above by $\alpha\lesssim 1$, since otherwise there is
no eternal inflation in this model.  We used the value $\alpha=1$.

The dependence of $N$ on $H_0$ is only logarithmic, but still the
number of e-foldings is too large for small values of $H_0$.  We used
$H_0=0.63$ (which corresponds to $V_0=0.05$).  This value of $H_0$ is
dangerously close to the Planck scale, but here we disregard all
complications of quantum gravity.  (Our choice of
parameters in these simulations
was dictated by the computational constraints, and we made no attempt
to make this choice realistic.)

In two-field simulations, the dispersion of $\chi$ due to quantum
fluctuations is
\beq
\delta\chi_q\sim{H_0\over{2\pi}}N^{1/2}\approx 0.6 \frac{H}{\sqrt{d}}.
\eeq
In order to have a representative simulation,
we require that $\chi$ should spread through its entire range
($\pi/\beta$) by the end of the simulation.  This
implies $\delta\chi_q\gtrsim \pi/\beta$, or $\beta\gtrsim 5d^{1/2}H^{-1}$.
In our simulations we used $d=1,~\beta=10$.

% Beginning of description of simulations

\subsection{Comoving space simulations}

The first series of simulations described the spacetime in co-moving
coordinates. The simulation was performed on a co-moving grid of points
in either $2$ or $1$-dimensional space. Both the one-field and the
two-field models were simulated. Initially the field values were taken
constant throughout the grid, with $\vp=0$.  At each timestep, the
fields were incremented by the classical evolution term \( \Delta
\varphi  \) from Eq.~(\ref{eq:dphic}) and
by the corresponding quantum fluctuation \( \delta \varphi \left( {\bf
x}\right) \).
The evolution of the field $\chi$ in the two-field model
was computed in the same manner.
The simulation  was continued until the physical distance
between adjacent grid points grew larger than the horizon
size.\footnote{In the ($2+1$)-dimensional case, we extended the
simulations somewhat beyond this point and allowed the separation of
adjacent grid points to exceed the horizon in the inflating regions with
$\vp$ near the top of the potential.  This should not have had any
effect on the thermalized regions shown in Figures 2 and 5.}  The 
expansion factor was calculated for each point to determine the
probability  distribution of $\chi$ weighted by the physical
volume.

Instead of taking the sine waves,  we used for \( \delta \varphi \left(
{\bf x}\right) \) true Gaussian random fields with an appropriate
correlation function. To obtain the increments for the grid points at
each step, one only needs to generate a set of Gaussian variables \(
\delta \varphi \left( {\bf x}_i\right) \) with zero mean and correlation
matrix
\beq
\langle \delta \varphi \left( {\bf x}_i\right) \delta \varphi \left(
{\bf x}_j \right) \rangle \equiv C_{ij}=C\left( d\left( {\bf x}_i,
{\bf x}_j \right) \right),
\eeq
where $d$ is the geodesic distance between points \( {\bf x}_i \), \(
{\bf x}_j \) calculated using the known scale factors at intermediate
points. Given the symmetric and positive-definite correlation matrix
$C_{ij}$, one can compute its Cholesky decomposition \( C = L L^T \)
where $L$ is a lower triangular matrix (see e.g.~Ref.~\cite{NR} for a
numerical method), and the required vector of random increments is then
obtained as
\beq
\delta \varphi \left( {\bf x}_i\right) = L_i^j u_j
\label{eq:dphigau}
\eeq
where $u$ is a vector of uncorrelated random numbers with normal
distribution. The correlation function \( C(r) \) decays rapidly over
the distance of a few horizon sizes and was taken to be zero at
distances larger than \( 2 H^{-1} \) to simplify calculations.

We have performed simulations for the double-well model (\ref{v}) and
the two-field model (\ref{vphichi}).  Momentary snapshots and spacetime
slices of some of these simulations are shown in Section II.  A more
systematic investigation of the spacetime structure of eternally
inflating universes, based on numerical simulations of this sort, will
be given in a separate publication.

\subsection{Physical space simulations}

%The simulation methods outlined in the preceding section can
%be used to calculate the probability distribution ${\cal
%P}_*(\chi)$.
In the second series of simulations we have investigated the
thermalization surface of
%We performed this calculation for
the two-field model of Eq.~(\ref{vphichi})
with the parameter values (\ref{ex3})
%\beq
%\alpha=1,~~\beta=10,~~\lambda=0.1,~~H_0=0.63.
%\label{parameters}
%\eeq
in $(1+1)$ dimensions.  We have already
discussed the spacetime structure of the universe in this model
(Section IIB).  As illustrated in Fig.~\ref{fig:6}, the universe
contains an infinite number of
disconnected, infinite, spacelike thermalization lines (thermalization
surfaces are lines in a $(1+1)$-dimensional simulation).  In order
to implement our prescription for the calculation of probabilities, we
have to choose one of these lines and calculate ${\cal
P}_*(\chi)\propto d\ell/d\chi$ on
a large segment of that line.  Here, $d\ell$ is
the length of all parts of the segment where $\chi$ takes values in
the interval $d\chi$.  A segment of length $2R$ is a
one-dimensional analogue of a sphere of radius $R$.  The probability
distribution is obtained in the limit of large $R$, so we used the
whole length of the thermalization line that we were able to generate.

The simulation space is a line of gridpoints, each carrying a
value of \( \varphi \) and \( \chi \); initially the values of the
fields are set constant everywhere. At every timestep, the fields
are evolved by adding the classical drift \( \Delta \varphi \) and
the quantum fluctuation \( \delta \varphi \left( x\right) \)
generated from Eq.~(\ref{sin}) (we chose this over the more
precise Eq.~(\ref{eq:dphigau}) for reasons of computational
feasibility) in which the sine wave was multiplied by a random
factor \( M \),
\beq
\delta\varphi({\bf x})=M{H\over{\sqrt{2}\pi}}\sqrt{H\Delta \tau}
\sin\left( He^{H\tau}({\bf n\cdot x}+\alpha)\right).
\label{eq:Msin}
\eeq
It can be verified that if \( \alpha \) is uniform in \( \left[
0,2\pi \right] \) and \( M \) has the standard \( \chi ^2 \)
distribution with two degrees of freedom, then \( M\sin \alpha \)
has normal distribution. This modification of Eq.~(\ref{sin})
circumvents the problem we mentioned above with the unphysical
distribution of the fluctuation amplitude resulting from a
fixed-amplitude sine wave.

The local expansion factor is also
computed at 
each point. Due to the one-dimensional geometry of the simulation space, it is
possible to maintain an (approximately) constant number of points per physical
horizon size by inserting new grid points as necessary at each step. Thus we
obtain a connected piece of the thermalization line in \emph{physical} (instead
of co-moving) coordinates. 

Since the calculation involves only a single thermalization line, one
can save a great deal of memory and computer time by discarding parts
of the simulation that do not affect that particular line.  At any
given moment the universe in the simulation consists of thermalized
regions of two types (corresponding to $\varphi=\pm\eta$) separated by
inflating regions.  Suppose we have chosen a plus-region (with
$\varphi=+\eta$) for the calculation of ${\cal P}_*(\chi)$.  This
region will never merge with any of the minus-regions, so we can
safely discard the nearest minus-regions on both sides and everything
beyond them.  New minus-regions can later be formed in the remaining
part of the simulation, and we can discard them and everything beyond
them again.  A greater efficiency is achieved by discarding the
nearest thermalized regions regardless of their type, so the
simulation includes only one thermalized region and two adjacent
inflating regions.  Once again, the procedure is repeated whenever new
thermalized regions are formed.  If the discarded nearest neighbors were
plus-regions, they can later merge with our region of choice, in which
case the simulation will terminate prematurely.  This can be fixed by
going back to the moment when the merged neighboring region was
discarded and
performing the simulation for that region from that time on.  The
resulting thermalization line is then attached to the line obtained
for the initial region.  We can make an arbitrary number of such
attachments to generate a thermalization line of any desired length.
We used this attachment procedure in our simulations.

\begin{figure}
\psfig{file=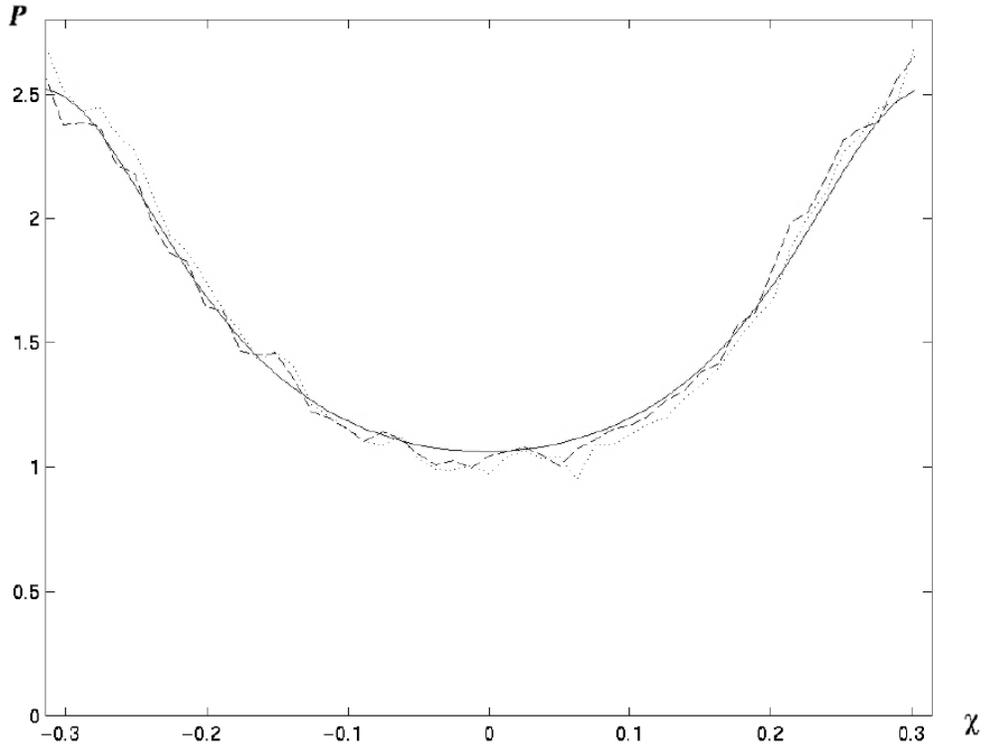,width=6in,angle=0}
\caption{The probability distribution for $\chi$ for the model
parameters (\ref{ex3}).  The distributions obtained directly from two
different thermalization lines in a simulation are shown by dotted and
dashed lines.  The solid line shows the distribution obtained by
solving the FP  equation.}
\label{fig:13}
\end{figure}

Fig.~\ref{fig:13} shows the probability distributions we obtained for two
different thermalization lines in our simulations.
As expected, the two distributions
are very similar.
Also shown is the distribution we obtained by solving
the Fokker-Planck equation in Section IV C.  
It agrees with the simulation-based
distributions with an accuracy of 4\% .

\section{Conclusions}

In this paper we analyzed the origin of the gauge dependence of
probability distributions for observables in models of eternal
inflation.  We reviewed the resolution
of this problem proposed in Ref. \cite{AV98}, gave a more precise
formulation of this proposal and used it to develop two methods
for calculating probabilities.

The root of the problem is that an eternally inflating universe
contains an infinite number of observers (galaxies), and one needs to
introduce some cutoff procedure in order to calculate
probabilities.  The proposed procedure can be summarized as follows.
Pick an arbitrary galaxy ${\cal G}$ at an arbitrary moment in its
history; this defines a point in spacetime.
Consider the hypersurface of constant energy density $\Sigma$ passing
through that
point.  Define a sphere of radius $R$ to include all points
whose geodesic distance from ${\cal G}$ along $\Sigma$ is less than
$R$.  The proposal is to evaluate the probability distribution
$P(\chi)$ based on
galaxies formed within the co-moving region defined by a sphere of
radius $R_c$ and then let $R_c\to\infty$.  The surface $\Sigma$ is an
infinite spacelike surface, and the distribution of the fields
$\chi_a$ on this surface is statistically homogeneous on large scales.
One expects, therefore, that the
resulting probability distribution $P(\chi)$
will be independent of the choice of the
``central'' galaxy
${\cal G}$ and that it will rapidly converge once the cutoff radius
$R_c$ becomes larger than the characteristic scale $\xi$ of variation of the
fields $\chi_a$ on $\Sigma$.

One argument in favor of this proposal is that it satisfies the
``correspondence principle'' with models of
finite inflation, where the definition of probabilities is
unambiguous.  In such models, the total number of galaxies in the
universe is finite, and all galaxies should in principle be included
in the calculation of $P(\chi)$.  However, a spherical cutoff with
$R_c\gg\xi$ should give a distribution very close to that obtained
using the complete set of galaxies.

We performed numerical simulations of eternal inflation in a two-field
model (\ref{vphichi}) and showed how the spherical cutoff can be
implemented to calculate the probability distribution $P(\chi)$
numerically.
Due to significant computational constraints, we were able to perform
simulations only for a rather restricted set of parameter values.

An alternative approach, which does not suffer from these restrictions,
is to calculate $P(\chi)$ by solving the Fokker-Planck equation with
the inflaton field $\vp$ playing the role of time.  This form of the
FP equation is valid only in the slow-roll range, $\vp_q\lesssim
\vp\lesssim \vp_*$, and the method can be used only in cases where one
can specify the initial distribution $P(\vp_0,\chi)$ at some
$\vp_0\gtrsim\vp_q$.  We have shown in Section IV that this can be done in a
reasonably wide class of models.  In more general models, one may have
to do numerical simulations in the quantum diffusion regime in order
to determine the initial conditions for
the FP equation in the slow-roll regime.

We have indicated some cases in which the FP equation can be
approximately solved analytically.  In one of these cases, when the
fields $\chi_a$ remain nearly constant during the slow roll of $\vp$,
the resulting distribution agrees with that obtained in \cite{AV98}.
We obtained a numerical solution of the FP equation for the parameter
values used in our simulations and verified that the probability
distribution $P(\chi)$ obtained in this way agrees with the
distribution calculated directly from the simulations.

The approach we developed in this
paper can be extended to models of ``chaotic'' inflation, where the
inflaton potential rises to super-Planckian values.  However, in such
models one has to deal with the uncertainties associated with Planck
physics.  In particular, one has to impose some boundary condition for
the FP equation at the ``Planck boundary'' $\vp=\vp_p$, where
$V(\vp_p)\sim 1$.

Perhaps the most important application of our approach 
would be a probabilistic analysis of the density
perturbation spectra.  As we emphasized in the Introduction, this
problem arises in practically all inflationary models, with or without
the additional fields $\chi_a$.  A brief discussion of this issue can
be found in Ref.~\cite{AV98} where it was argued that 
the present approach should give essentially the same results as the
standard calculations \cite{14}.  A more detailed analysis will be
given elsewhere.

Our prescription for calculating
probabilities cannot be applied to models where $\chi$ is a discrete
variable.  In such models,
different values of $\chi$ are taken in different, causally
disconnected thermalized regions.
The spherical cutoff prescription
involves only a single thermalized region, and it is clear that it
cannot be used to determine the probability distribution in this case.
One can take this as indicating that no probability distribution for
a discrete variable can be meaningfully defined in an eternally
inflating universe.  Alternatively, one could try to introduce some
other cutoff prescription to be applied specifically in the case of a
discrete variable.  It is possible that some version of the
$\epsilon$-prescription of Ref.\cite{9} could be used for this
purpose.  This issue requires further investigation.

\section{Acknowledgments}

We are grateful to Jaume Garriga, Andrei Linde and Ken Olum for their comments on the manuscript. The work of A.V. was supported in part by the National Science Foundation. S.W. was supported by PPARC rolling grant GR/L21488.

\appendix

\section{Derivation of the reduced Fokker-Planck equation
in the limit of negligible diffusion}

We start from the usual time-dependent FP equation in two dimensions \(
\left( \phi ,\chi \right)  \) for the co-moving probability
distribution \( P\left( t,\phi ,\chi \right)  \), under assumption that
diffusion in \( \phi  \) direction is negligible and using the Ito
factor ordering:
\begin{equation} \label{tdde}
\partial _{t}P=\partial
^{2}_{\chi }\left( \tilde{D}\left( \phi ,\chi \right) P\right)
-\partial _{\chi }\left( \tilde{v}_{\chi }P\right) -\partial _{\phi
}\left( \tilde{v}_{\phi }P\right) .
\end{equation}
Here \(
\tilde{v}_{\phi } \) and \( \tilde{v}_{\chi } \) are the usual (both \(
\phi  \) and \( \chi  \)-dependent) drift coefficients,
\begin{equation}
\tilde{v}_{\phi }=-\frac{1}{4\pi }\partial _{\phi
}H,\quad \tilde{v}_{\chi }=-\frac{1}{4\pi }\partial _{\chi }H,
\end{equation}
and the diffusion coefficient is \( \tilde{D}=H^{3}/8\pi
^{2} \). The diffusion is bounded by the line \( \phi =\phi _{*} \) which is the
absorbing boundary (due to thermalization). No other exit
boundaries are present.

First we consider the probability density
\( P_{e}\left( t;\phi ,\chi ;\phi _{*},\chi _{*}\right) \)
to exit at a particular time \( t \) through the neighborhood of a
particular point \( \chi _{*} \) along the line \( \phi =\phi _{*} \), given that we started at \( t=0 \) at a
given point \( \left( \phi ,\chi \right) \). Obtaining the distribution \( P_{e}(t;...) \) is a standard problem in the
theory of stochastic processes \cite{StochasticLore};
once we know \( P_{e}(t;...) \), we can
integrate it over time until \( t=\infty  \) and obtain the total probability
to thermalize in the neighborhood of \( \chi _{*} \).

According to the
standard theory, we can find the \emph{generating function} \( g(s;...) \) for
the distribution \( P_{e}(t;...) \), defined as
\begin{equation} \label{gdef}
g\left( s;\phi ,\chi ;\phi _{*},\chi _{*}\right) \equiv \int
_{0}^{\infty }e^{-st}P_{e}\left( t;\phi ,\chi ;\phi _{*},\chi _{*}
\right) dt
\end{equation}
because it turns out that \( g \) is the
solution of the ``backward'' or adjoint diffusion equation
\begin{equation} \label{gequ}
\tilde{D}\left( \phi ,\chi \right)
\partial ^{2}_{\chi }g+\tilde{v}_{\chi }\partial _{\chi
}g+\tilde{v}_{\phi }\partial _{\phi }g=sg
\end{equation}
with boundary
condition at \( \phi =\phi _{*} \) given by
\begin{equation}
\label{gbc} g\left( s;\phi _{*},\chi ;\phi _{*},\chi _{*}\right)
=\delta \left( \chi -\chi _{*}\right) .
\end{equation}
This boundary
condition makes the function vanish everywhere on the exit boundary
except at the point \( \chi _{*} \) where we want to find the exit
probability. It is seen from Eq. (\ref{gdef}) that the total
probability of exit at any time, i.e.~the integral of \( P_{e} \)
through all \( t \), is equal to the value of \( g \) at \( s=0 \).
This value \( g\left( s=0;\phi ,\chi ;\phi _{*},\chi _{*}\right) \equiv
g_{0} \) is then interpreted as the probability that the diffusion
process finishes at \( \chi =\chi _{*} \) if it started at \( \left(
\phi ,\chi \right)  \). The boundary condition Eq.~(\ref{gbc}) means
that if we already started at the boundary but at a different \( \chi
\), then the probability to exit at the chosen value \( \chi =\chi _{*}
\) is zero. The equation for \( g_{0} \) is
\begin{equation}
\label{g0equ}
\tilde{D}\left( \phi ,\chi \right) \partial ^{2}_{\chi
}g_{0}+\tilde{v}_{\chi }\partial _{\chi }g_{0}=-\tilde{v}_{\phi
}\partial _{\phi }g_{0}.
\end{equation}
This is formally the same as
the backward Fokker-Planck equation for a one-dimensional diffusion
process with \( \phi  \) playing the role of time, where \( g_{0}\left(
\phi ,\chi ;\phi _{*},\chi _{*}\right)  \) is interpreted as the
probability of finishing at \( \chi =\chi _{*} \) at the final ``time''
\( \phi =\phi _{*} \) if started at \( \left( \phi ,\chi \right)  \).
The boundary conditions (\ref{gbc}) correspond exactly to that
interpretation. The drift and diffusion coefficients become
\begin{equation} v_{\chi }\equiv -\frac{\tilde{v}_{\chi
}}{\tilde{v}_{\phi }},\quad D\equiv -\frac{\tilde{D}}{\tilde{v}_{\phi
}} \end{equation} in agreement with
Eqs.~(\ref{eq:v-red})--(\ref{eq:D-red}).
The equation adjoint to Eq.~(\ref{g0equ})
is therefore interpreted as the diffusion equation for the
``time''-dependent probability distribution
\( P\left( \phi ,\chi ;\chi_{0}\right)  \),
with \( \phi  \) as ``time'' and the new kinetic
coefficients,
\begin{equation} \label{pequ}
\partial _{\phi }P=\partial
^{2}_{\chi }\left( DP\right) -\partial _{\chi }\left( v_{\chi }P\right)
.
\end{equation}

The equations for the physical volume distribution are derived in the
same manner and differ from Eqs.~(\ref{tdde})--(\ref{g0equ}) only by
the addition of the growth term \( 3HP \) or \( 3Hg \) as appropriate,
and the corresponding growth term in Eq.~(\ref{pequ}) is similarly
obtained as \( -3HP/\tilde{v}_{\phi } \). We therefore arrive at
Eq.~(\ref{FP}).

% Figures section: made automatically by RevTeX

% Draw only the bounding boxes of figures. Remove after debugging.
%\psdraft

\tightenlines

\end{document}